\documentclass[useAMS,usedcolumn,usenatbib]{mn2e}
\usepackage{epsfig}
\usepackage{graphicx}
\usepackage{psfrag}
\usepackage{url}
\usepackage{xspace}
\usepackage{color}

\newcommand{\be}{\begin{equation}}
\newcommand{\ee}{\end{equation}}
\newcommand{\bea}{\begin{eqnarray}}
\newcommand{\eea}{\end{eqnarray}}

\newcommand{\size}{R}
\newcommand{\bs}{}
\newcommand{\ellipticity}{\epsilon}

\def\bgamma{{\gamma}}

\def\eprinttmp@#1arXiv:#2 [#3]#4@{
\ifthenelse{\equal{#3}{x}}{\href{http://arxiv.org/abs/#1}{#1}
}{\href{http://arxiv.org/abs/#2}{arXiv:#2} [#3]}}
\providecommand{\eprint}[1]{\eprinttmp@#1arXiv: [x]@}
\newcommand{\adsurl}[1]{\href{#1}{ADS}}

\title[]{Cosmic shear requirements on the wavelength-dependence of telescope point spread functions}
\author[E. S. Cypriano et al.]{E. S. Cypriano$^{1,2}$, A. Amara$^{3}$,
 L. M. Voigt$^{2}$, S. L. Bridle$^{2}$, F. B. Abdalla$^{2}$,
 \newauthor
 A. R\'efr\'egier$^{4}$, M. Seiffert$^{5,6}$, J. Rhodes$^{5,6}$ \\
 \\
$^{1}$Departamento de Astronomia, Instituto de Astronomia, Geof\'{\i}sica e Ci\^encias Atmosf\'ericas, Universidade de S\~ao Paulo,
Rua do Mat\~ao 1226, Cidade Universit\'aria 05508-900, S\~ao Paulo, SP, Brazil\\
$^{2}$Department of Physics and Astronomy, University College London,
Gower Street, London, WC1E 6BT, UK\\
$^{3}$Department of Physics, ETH Z\"{u}rich, Wolfgang-Pauli-Strasse 16, CH-8093 Z\"{u}rich, Switzerland.\\
$^{4}$Service d\'Astrophysique, CEA Saclay, 91191 Gif sur Yvette, France \\
$^{5}$Jet Propulsion Laboratory, California Institute of Technology, 4800 Oak Grove Drive, Pasadena, CA 91109 \\
$^{6}$California Institute of Technology, 1201 E California Blvd., Pasadena, CA 91125, USA\\
}
\begin{document}

\date{Accepted . Received ; in original form }


\maketitle

\begin{abstract}
Cosmic shear requires high precision measurement of galaxy shapes in the
presence of the observational Point Spread Function (PSF) that smears out
the image. The PSF must therefore be known for each galaxy to a high
accuracy. However, for several reasons, the PSF is usually wavelength
dependent, therefore the differences between the spectral energy
distribution of the observed objects introduces further complexity.
In this paper we investigate the effect of the wavelength-dependence of
the PSF, focusing on instruments in which the PSF size is dominated by
the diffraction-limit of the telescope and which use broad-band filters
for shape measurement.

We first calculate biases on cosmological parameter estimation from
cosmic shear when the stellar PSF is used uncorrected. Using realistic
galaxy and star spectral energy distributions and populations and a
simple three-component circular PSF we find that the colour-dependence
must be taken into account for the next generation of telescopes. We
then  consider two different methods for removing the effect (i) the use
of stars of the same colour as the galaxies and (ii) estimation of the
galaxy spectral energy distribution using multiple colours and using a
telescope model for the PSF. We find that both of these methods  correct
the effect to levels below the tolerances required for per-cent level
measurements of dark energy parameters. Comparison of the two methods
favours the template-fitting method because its efficiency is less 
dependent on galaxy redshift than  the broad-band colour method and takes
full advantage of deeper photometry.

\end{abstract}


\section{Introduction}

Measurements of the cosmic shear signal are expected to play a leading role in
furthering our understanding of our Universe, in particular the nature of dark
matter and dark energy or its possible alternatives such as modifications in
gravity. The gravitational lensing of light from distant galaxies by intervening
mass provides a powerful insight into the growth of structure and the expansion
history of the universe \citep[for recent reviews
see][]{HJRev08,Munshi08,Refregier:2003ct}.

Several planned future dark energy missions are designed with weak
lensing as a primary science driver, including ground-based
projects: the KIlo-Degree Survey (KIDS)~\footnote{\url{http://www.astro-wise.org/projects/KIDS}}; the Panoramic Survey
Telescope and Rapid Response System
(Pan-STARRS)~\footnote{\url{http://pan-starrs.ifa.hawaii.edu}}; the
Dark Energy Survey
(DES)~\footnote{\url{http://www.darkenergysurvey.org}}; and the
Large Synoptic Survey Telescope
(LSST)~\footnote{\url{http://www.lsst.org}}, and space missions
Euclid~\footnote{\url{http://www.euclid-imaging.net}}
\citep{euclid_assesstudy,euclid_scibook}
and the Joint Dark
Energy Mission (JDEM)~\footnote{\url{http://jdem.gsfc.nasa.gov}}.
The success of the method relies on accurate measurement of galaxy
shapes \citep[e.g.][]{STEP1,STEP2,great08}.

As light from galaxies passes through the atmosphere, telescope
optics and measurement devices it is convolved with a kernel,
referred to as the Point Spread Function (PSF). The size of the PSF
is similar to the size of the galaxies used to measure cosmic shear,
making accurate determination of the underlying galaxy shape a
significant challenge \citep[e.g.][]{lewis09,voigt&bridle09}.
Precise modelling of the PSF is
crucial~\citep[e.g.][]{paulinavrb08}. This can in principle be
performed using accurate knowledge of the telescope optics, however
in practice it is usual to make use of the point-like nature of
stars by modelling the PSF based on the shapes of stellar images.
The situation is complicated by (i) variations in the shape of the
PSF across the detector and (ii) the wavelength-dependence of the
PSF shape. In this paper we concentrate on the second of these two
effects, focusing on a telescope which is nearly
diffraction limited and on a instrument that uses broad-band optical
filters for shape measurement.

\begin{figure}
\centerline{ \includegraphics[width=1.0\columnwidth]{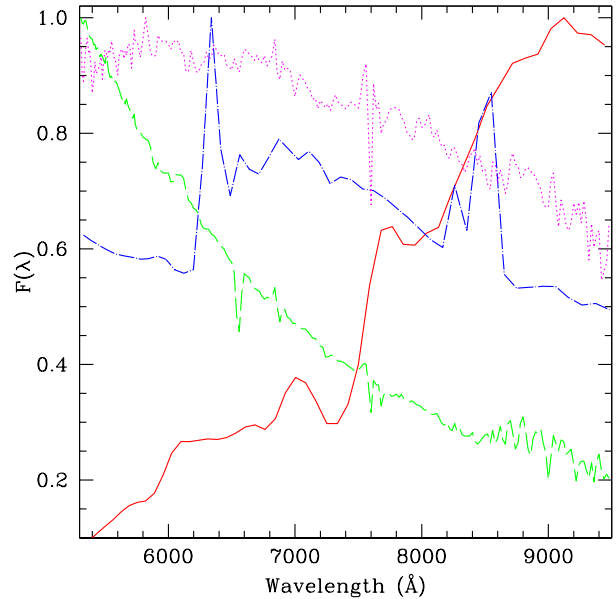}}
\caption{A sample of stellar and galaxy spectra showing the variety
of SEDs among astronomical objects. An elliptical galaxy at $z=0.9$ (solid red),
an irregular galaxy
at $z=0.7$ (dot-dashed blue), an A0V star (dashed green)
and a K0III star (dotted magenta). \label{fig:spectra}}
\end{figure}

Stars and galaxies have different spectral energy distributions
(SEDs), both intrinsic and observed, specially if we
consider that galaxies are observed over a broad range of redshifts
(See Figure \ref{fig:spectra}). Therefore the PSF shape measured
from a stellar image will be different from the true PSF applied to
the galaxy, leading to a bias on the measured underlying galaxy
shape. The wavelength-dependence of the PSF has a greater impact if
observations are performed using broad-band filters.
 In this paper we quantify the effect of ignoring the wavelength-dependence of the PSF in a Euclid-like space mission and show the
feasibility of reducing the effect by using colour information. We
determine the wavelength-dependence of the PSF shape using a simple
model for the convolution kernel and applying realistic galaxy and
stellar spectral energy distributions (SEDs).

For the reasons we will discuss in the following sections this
paper will focus on the wavelength-dependence of the PSF size. There are several
measures of this quantity, such as the 50
we adopt the full width at half-maximum intensity (FWHM) as our size parameter
given that this is a more familiar quantity to the astronomical community and
describes well the simple PSF models we use in this paper.

The paper is organised as follows: in Section~\ref{sec:propagation}
we set out a formalism for propagating a mis-estimation of the PSF
shape through to biases on cosmological parameters. In
Section~\ref{sect:psf_model} we describe the separate contributions
to the PSF model and in Section~\ref{sect:impact} we predict the
range in PSF sizes expected for two different broad-band filters
using a realistic distribution of stellar and galaxy SEDs and
 investigate two different
methods for reducing the bias. These involve using: (i) galaxy and
stellar colours to correct the PSF size
and (ii) template fitting. Also we place requirements on a simple
model for the wavelength dependence for a Euclid-like survey.
Finally, in Section \ref{sec:discussion} we discuss our findings.

\section{Implications of using an incorrect PSF}
\label{sec:propagation}

We first make a simple calculation of the shear bias induced by
mis-estimating the size and ellipticity of the PSF and compare it to
the general systematics limit calculated in \cite{amarar07}. We then
consider the redshift dependence of the shear biases and propagate
this through to biases on cosmological parameters.

The shear measurement bias caused by an incorrect PSF will depend on
the shear measurement method employed.
For simplicity we assume here that shears are calculated
using unweighted quadrupole moments. Although this method is not
feasible in practice due to the low signal-to-noise level in real
images, it is related to the widely used Kaiser, Squires and
Broadhurst method~\citep{kaisersb95}. It has the advantage of being
extremely easy to use for shear measurement bias calculations.

Shear mis-estimates are often quantified by Taylor expanding the
estimated two-component shear $\hat{\gamma}_i$ ($i={1,2}$) in terms
of the true shear $\gamma_i$ as \be \hat{\gamma_i}= m_i \gamma_i +\
c_i \label{eq:mcdef}
\ee where $m_i$ is referred to as the multiplicative bias, $c_i$ as
the additive bias \citep{STEP1} and it is often assumed $m_1 \simeq
m_2 \equiv m$ and $c_1 \simeq c_2 \equiv c$.

\cite{amarar07} showed that, for a full-sky survey ($2 \times 10^4$ square degrees of
extragalactic sky) with 35 galaxies per square arcminute and a median redshift of 0.9, the shear
multiplicative error $m$ must be $<1\times 10^{-3}$ to keep systematic biases on cosmological
parameters below random uncertainties, for a range of possible redshift evolution scenarios for
$m$. Using the equations in Appendix~\ref{sec:mc_quadmoms} this translates to a requirement on
the PSF size mis-estimate $\delta F_{\rm PSF}$ of
\bea
\frac{\delta F_{\rm PSF}}{F_{\rm PSF}} \simeq m  \le  1 \times 10^{-3}.
\label{eq:deltaF_to_m}
\eea
\cite{amarar07} also placed a requirement on the mean square error $\sigma_{\rm sys}^2 < 10^{-7}$
for the same survey. Interpreting this as a requirement on the square of the shear measurement
additive error $c^2$, and neglecting the subdominant term in Eq.~\ref{eq:c}, this places a
requirement on the PSF ellipticity mis-estimate $\delta \ellipticity_{{\rm PSF}\, i}$ of
\bea
\delta \ellipticity_{{\rm PSF}\, i} \simeq 4 c \le 1 \times 10^{-3} .
\eea

Because the observed SED of a galaxy changes with redshift, the appropriate
PSF will also depend on galaxy redshift. Therefore the PSF biases $\delta F_{\rm PSF}$
and $\delta \ellipticity_{\rm PSF}$ depend on redshift and so do the multiplicative
and additive biases $m$ and $c$.
If, for example, the impact of the multiplicative and additive biases as a
function of redshift mimics a particular cosmological parameter the
requirements may be more stringent than the above more approximate
calculation. We will therefore calculate PSF biases as a function of galaxy
redshift and insert them into the more detailed calculation described below,
which propagates the effect into biases on cosmological parameters.

Use of the wrong PSF model will cause the measured cosmic shear
cross power spectra between redshift bins $i$ and $j$,
$\hat{C}^{\kappa}_{ij}(\ell)$, to differ from the true cosmic shear
power spectra $C^{\kappa}_{ij}(\ell)$. If a particular systematic on
the cosmic shear power spectrum $\Delta C^{\kappa}_{ij}(\ell) =
\hat{C}^{\kappa}_{ij}(\ell)-C^{\kappa}_{ij}(\ell)$ is ignored then
the bias on cosmological parameters $\delta p_{\alpha}$ is given by
\be \delta p_{\alpha} = \mathcal{F}_{\alpha \beta}^{-1}
\,\,\sum_{\ell} \,\,\Delta C^{\kappa}_{ij} \,\,\,\left({\rm
Cov}\left[C^{\kappa}_{ij}(\ell),C^{\kappa}_{kl}(\ell)\right]\right)^{-1}
\,\, \frac{\partial C^{\kappa}_{kl}(\ell)}{\partial p_{\beta}} \ee
where $i$, $j$, $k$, $l$ and $\beta$ are summed over, ${\rm
Cov}\left[C^{\kappa}_{ij}(\ell),C^{\kappa}_{kl}(\ell)\right]$ is the
two dimensional covariance matrix between the cross-spectra and
$\mathcal{F}$ is the Fisher matrix between the cosmological
parameters \citep{huterertbj06,amarar07}. 

In the presence of a redshift dependent multiplicative bias, the
measured lensing power spectrum can be given in terms of the true
lensing power spectrum by \be \hat{C}^{\kappa}_{ij}(\ell) =
C^{\kappa}_{ij}(\ell) (1 +\ m^i +\ m^j) \ee where $m^i$ is the
multiplicative bias for redshift bin $i$, averaged over all galaxies
\citep[][Eq. 16]{huterertbj06}.

The impact of additive errors depends to first order on the spatial
variation of the additive errors. For the case of a wavelength
dependent PSF this will induce power on the scale of the separation
between stars of a typical colour, which may be propagated into
cosmology \citep[see also][]{Guzik05}.
Here we focus on PSF size mis-estimates and therefore do not
consider additive errors further.

In this paper we compare the PSF sizes for stars with those for
galaxies, without considering the galaxy morphology or profile. The
equations derived above and in the Appendix make it possible to draw
significant conclusions about the cosmology biases independent of
considerations about the galaxy light distribution, if all parts of
the galaxy have the same colour. Our main metric is the difference
between the FWHM of the stellar and galaxy PSFs. We average this
over populations of galaxies for various different PSF correction
schemes.

\section{The PSF model}
\label{sect:psf_model}

\begin{figure}
\centerline{ \includegraphics[width=1.0\columnwidth]{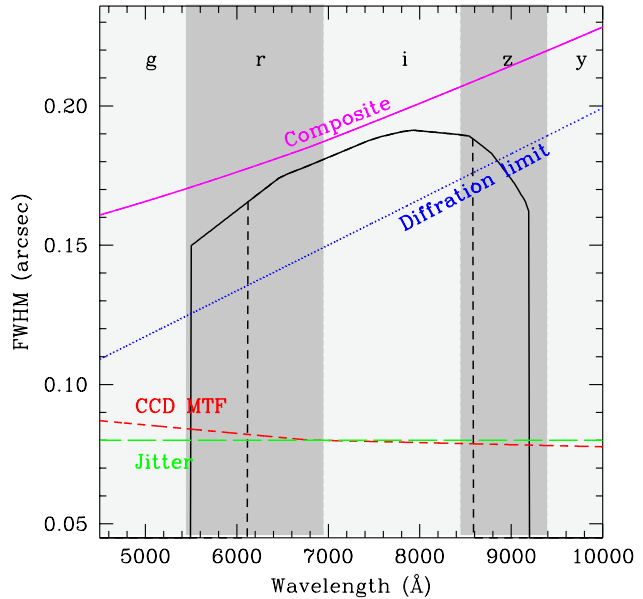}}
\caption{\label{coldep}
Contribution from each compoment of the PSF to
the image quality (FWHM) as a function of wavelength.
The dotted (blue)
line represents the diffraction component, the long-dashed (red)
line the CCD modulation transfer function component, the short-dashed
(green) the achromatic component and the thick solid line (magenta)
shows the overall result taking into account all the previously
described components. The solid (black) line shows the relative
instrumental passband using a wide optical filter $F1$
such as that proposed for Euclid. The dashed (black) vertical lines
show an alternative, narrower, lensing filter ($F4$). The background shaded areas show approximately the wavelength coverage of the $grizy$ filters.}
\end{figure}

We consider a simple instrument model in which the PSF is made up of three
circular components, each with a different wavelength-dependence. This is
reasonable for a space-based instrument composed mainly of reflective surfaces,
such as Euclid. For such an instrument the PSF ellipticity is relatively
insensitive to wavelength and the three main PSF components are (i) nearly
diffraction limited telescope optics giving rise to an Airy disk with size
inversely proportional to the wavelength (ii) the CCD modulation transfer
function (MTF) which tends to spread out higher energy photons more than lower
energy photons and (iii) a wavelength independent part such as telescope jitter.
We assume for simplicity that each component is Gaussian with a
wavelength-dependent size. 

We describe the total size of the PSF  by its FWHM, $F$, which  is given by the
quadratic sum of the FWHM values of the three components 
\begin{equation}
F_{\rm PSF}^2(\lambda)=F_{\rm D}^2(\lambda) + F_{\rm MTF}^2(\lambda) + F_{\rm J}^2
\label{eqn:fwhmcont}
\end{equation}
Note that the addition in quadrature works reasonably well even
if the diffraction limited component is not a Gaussian: we find adding FWHMs in
quadrature works to better than 5 per cent accuracy when an Airy disk is
convolved with a Gaussian of the same FWHM, and improves to better than 2 per
cent accuracy if the ratio of FWHMs is changed by a factor of 4 either way.

The  size of the  diffraction  limited image  is given by
\begin{equation}
F_{\rm D}(\lambda) = 0.154"~\left(\frac{D}{1.2 \rm m}\right)^{-1}~\left(\frac{\lambda}{7350 \rm \AA}\right)
\end{equation}
where  $D$ is  the diameter of the  primary mirror.
We take  the contribution from the CCD MTF to be that  measured empirically for
an e2v CCD 231-84 ({M. Cropper, priv. comm.}),  which is given approximately by 
\begin{equation}
F_{\rm MTF}(\lambda) = \left\{ \begin{array}{rl}
0.11" -0.027" \left(\frac{\lambda}{7000 \rm \AA}\right) &\mbox{ if $\lambda \le 7000\rm \AA$} \\
0.09" -0.007" \left(\frac{\lambda}{7000\rm \AA}\right)  &\mbox{ if $\lambda > 7000\AA$}
\end{array} \right.
\end{equation}
Finally, we take  the contribution to the PSF size from the achromatic component
to be
\begin{equation}
\mathrm{FWHM}_{\rm J} = 0.08"
\end{equation}
as appropriate for a Euclid-like instrument. 

We plot the three contributions to the PSF image size in
Fig.~\ref{coldep} for a 1.2 meter primary mirror. The image size is
dominated by the diffraction limit of the instrument.
Assuming the PSF contributions from different wavelengths all have
the same centroid we can calculate the FWHM of the composite PSF
from the FWHM of each component and the transmitted flux
$S(\lambda)T(\lambda)$, where $S(\lambda)$ is the spectral energy
distribution (SED) of the object and $T(\lambda)$ is the
instrumental plus filter response, by
\begin{equation}
\label{eq:final}
F_{\rm PSF}^2 =  {\int{S(\lambda)\; T(\lambda)\; F_{\rm PSF}^2(\lambda)~d\lambda}
               \over \int{S(\lambda)\; T(\lambda)~d\lambda}}.
\end{equation}
In this paper  we assume $T(\lambda)$ is the instrumental response for the case where
a wide optical F1 filter (5500--9200\rm \AA; see Fig.~\ref{coldep}) is used
for measuring the shapes of galaxies,  as proposed for the Euclid satellite.

\section{PSF and cosmology biases}
\label{sect:impact}

\begin{figure}
\centerline{ \includegraphics[width=1.0\columnwidth]{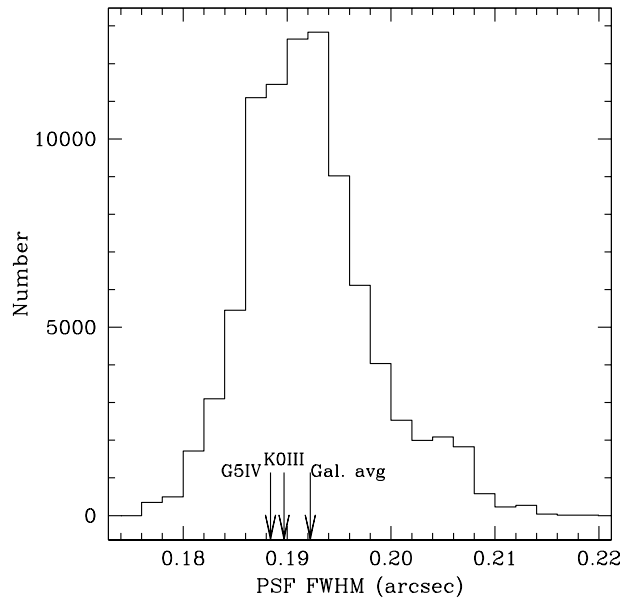}}
\caption{\label{fig:histpsf}
Histogram of the predicted PSF sizes (FWHM) for
the mock galaxy catalogue.
The average value of the distribution is marked on the
plot with an arrow, as well as the sizes for typical disk and halo stars G5IV and K0III.
}
\end{figure}

We use realistic galaxy and stellar populations to quantify the
amount by which the PSF FWHM is likely to be mis-estimated. Galaxy
SEDs are generated using mock catalogues designed to simulate the
distribution of redshifts, colours and magnitudes of galaxies in
GOODS-N \citep{Cowie04,Wirth04}. Template spectra are taken from
\citet{cww80} and \citet{kinney96} and intermediate types obtained
by linear interpolation of these templates \citep[for further
details see][]{abdallaea07}.
In this paper we include galaxies up to a redshift of 2.
Galaxies more distant than this will have small apparent sizes and
are thus unlikely to be used to measure cosmic shear. For instance, for
the Cosmos survey, much less than 10\% of the lensing usable galaxies have $z>2.0$
\citep{cosmos}.
The PSF sizes for stars are estimated using stellar SEDs from the
Bruzual-Persson-Gunn-Stryker (BPGS) Spectrophotometric
Atlas\footnote{\texttt http://www.stsci.edu/hst/observatory/cdbs/
astronomical\_catalogs.html\#bruzual-persson-gunn-stryker.}. The
catalogue contains 175 different SEDs covering a broad range of
spectral types. The true PSF size for each of the star and galaxy types in the
mock catalogues is then estimated by inserting its SED into Eq.~\ref{eq:final}.

A histogram of the FWHM  distribution for the galaxy population is
shown in Fig.~\ref{fig:histpsf}. The PSF size ranges from
approximately 0.175" to 0.220", has an average of
0.1922" and a dispersion of 0.006", or 3\% of the mean PSF size.
As an example, the FWHM of a G5 sub-giant star
(taken here as a typical disk star) is typically 0.1884 arcsec,
whereas a K0 giant star (typical halo/bulge star) is typically
0.1897 arcsec. In a conventional analysis which ignores the
wavelength-dependence of the PSF, galaxies with small angular
separations from the above  two example stars will often have their shears
underestimated. In fact,  out of all the stars in the BPGS catalogue only a
quarter of them (cold K and M stars)  have PSF estimated-sizes larger than the
average of the galaxies. The multiplicative shear mis-estimates for the example
G5 and K0 stars   are of the order of $m\sim3\times 10^{-2}$ relative to the
average galaxy,  in clear disagreement with our requirements on the PSF size
error (Eq. \ref{eq:mcdef}) and therefore we need to explore  methods for
mitigating the effect.
Below we describe two methods for correcting the PSF
wavelength-dependence using colour information.

Fluxes are obtained for each object in the mock catalogues for the filters $F1$,
$Y$, $J$ and $H$ up to the limiting magnitudes 26.25, 24.0, 24.0 and 24.0 respectively (AB
magnitudes, $5\sigma$ detections). In addition, we consider different scenarios
for the complementary ground based photometry. We assume here observations in
the filters  $g,r,i,z$ and $y$ with three different depths each,
shallow, medium and deep (see Table~\ref{table:gbphots}).
The fiducial optical depth used in this paper (medium) is chosen
to correspond to a DES or Pan-STARRS type survey with two dedicated telescopes
(PS2). Shallow and deep correspond to a Pan-STARRS type survey with one (PS1) or
four (PS4) dedicated telecopes.

\begin{table}
\caption{\label{table:gbphots} Pan-STARRS (PS1, PS2 and PS4) and DES optical photometry depths. The values
quoted here correspond to AB magnitudes of 5$\sigma$ detections.}
\begin{center}
\begin{tabular}{|l|cccc|}
\hline
Band & PS1   & PS2 & PS4 & DES\\
\hline
\hline
$g$ & 24.66 & 25.53 & 26.10 & 25.35\\
$r$ & 24.11 & 24.96 & 25.80 & 24.85\\
$i$ & 24.00 & 24.80 & 25.60 & 25.05\\
$z$ & 22.98 & 23.54 & 24.10 & 24.65\\
$y$ & 21.52 & 22.01 & 22.50 & 22.15\\
\hline
\end{tabular}
\end{center}
\end{table}

\begin{figure*}
\begin{tabular}{cc}
\includegraphics[width=1.0\columnwidth]{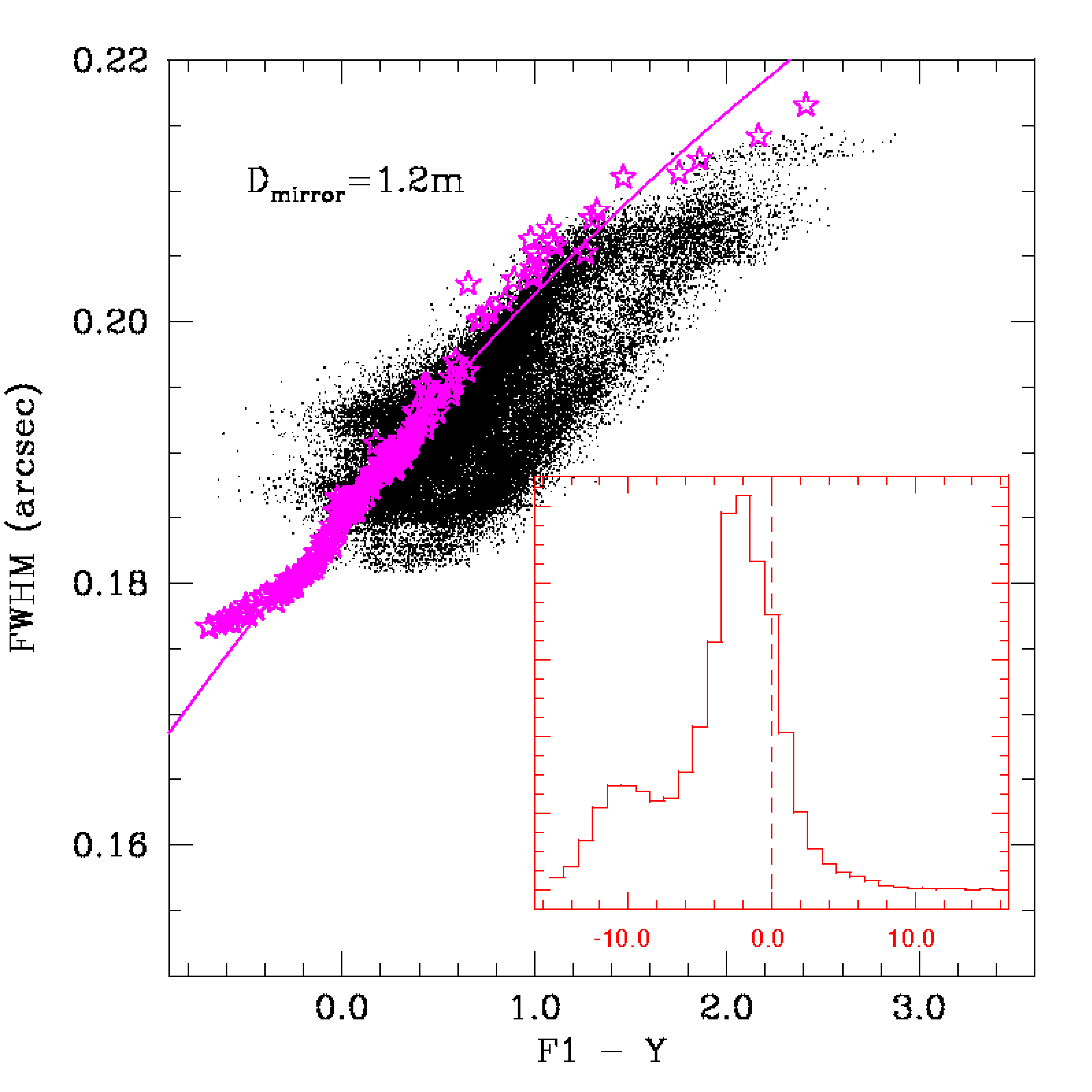} &
\includegraphics[width=1.0\columnwidth]{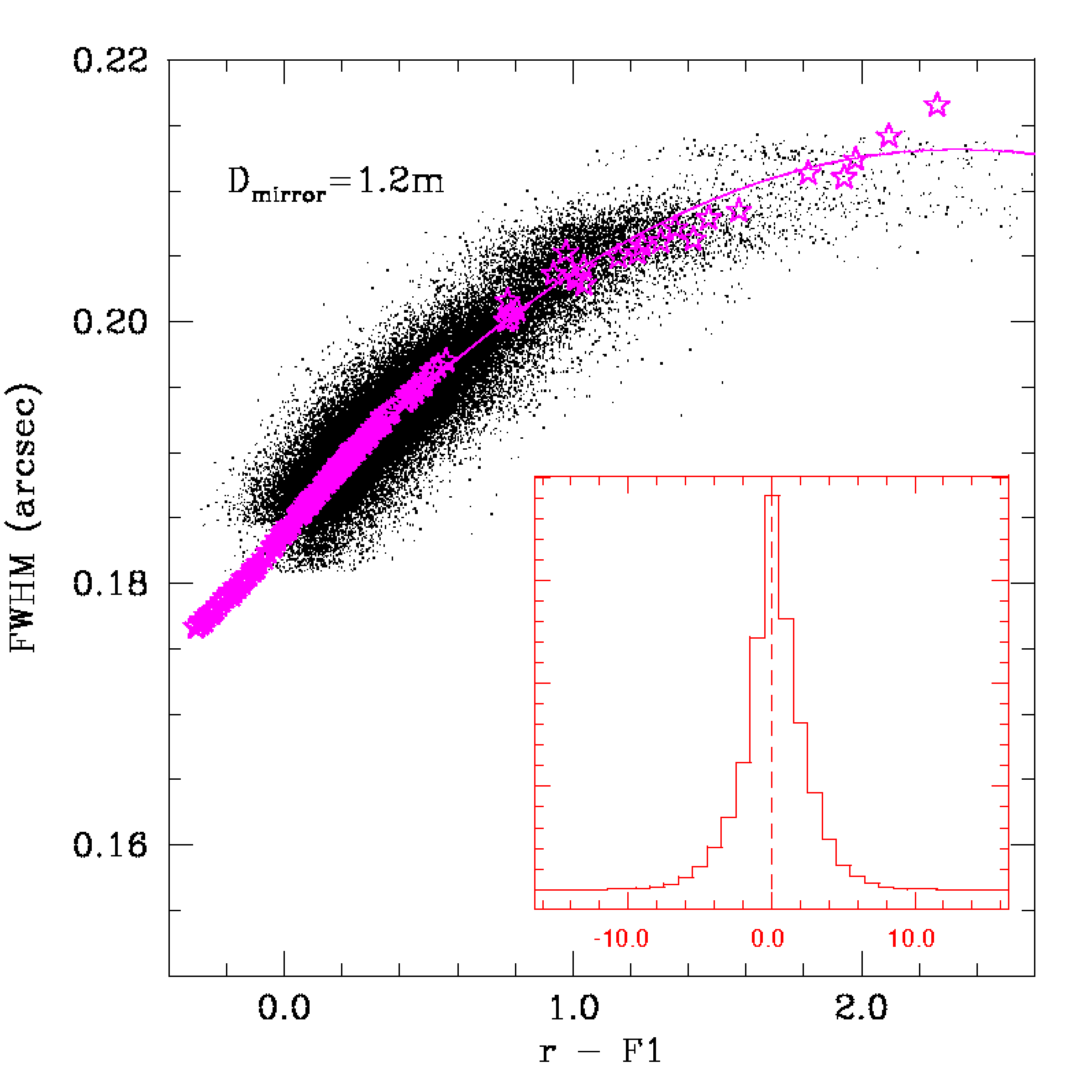} \\
\includegraphics[width=1.0\columnwidth]{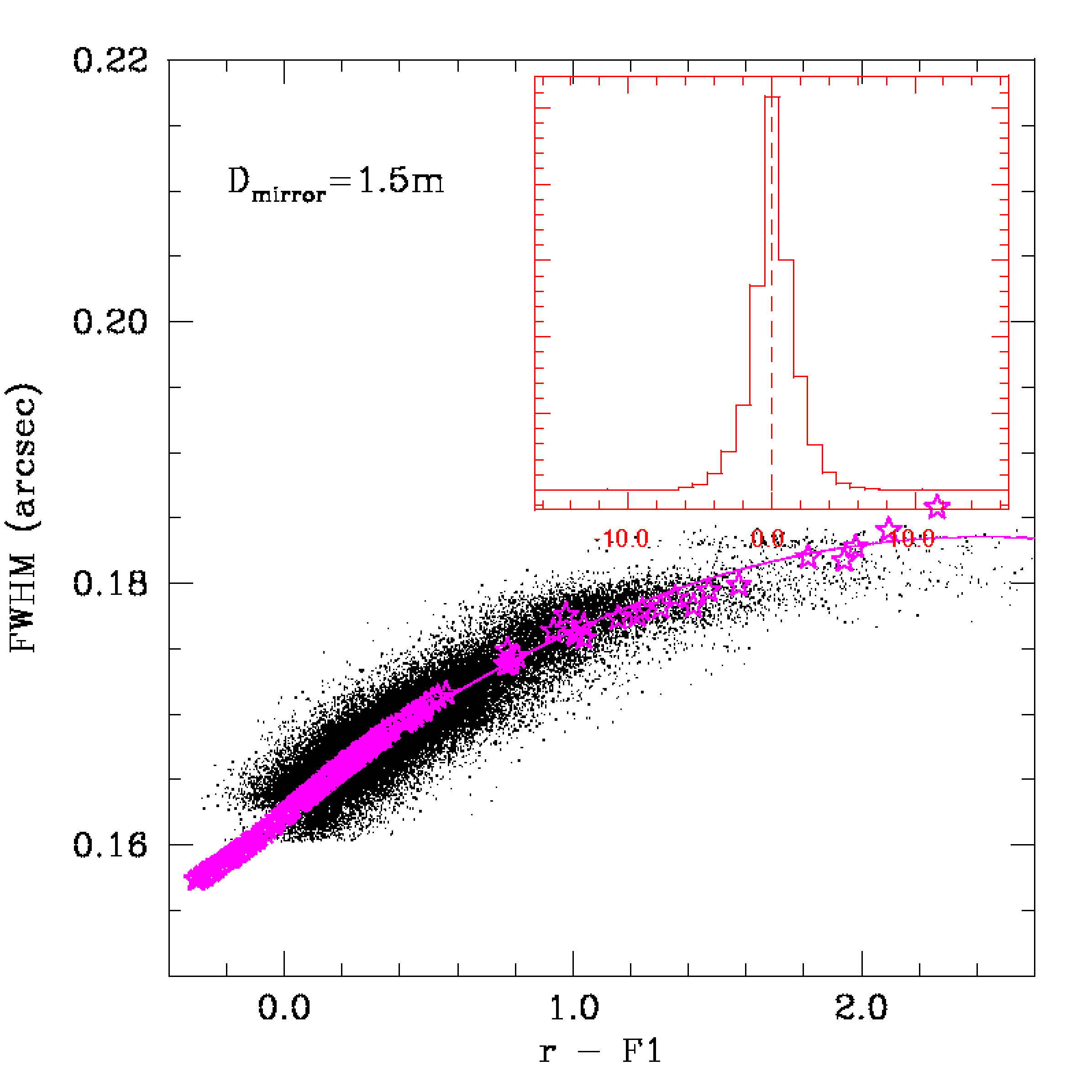} &
\includegraphics[width=1.0\columnwidth]{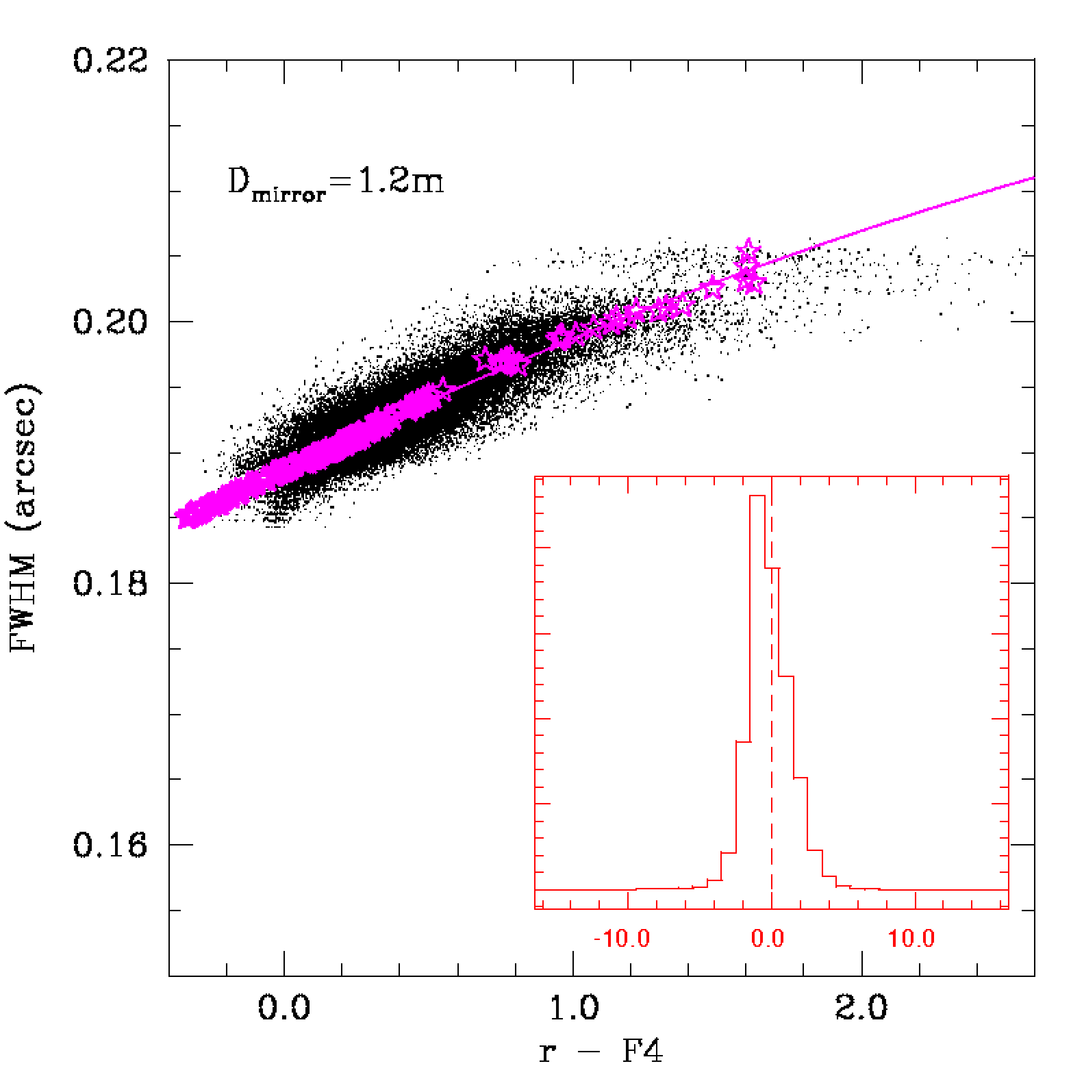}
\end{tabular}
\caption{\label{fig:FWHM-colour} PSF FWHM versus colour for a
realistic distribution of galaxies (black dots) and stars (magenta stars) of several
different spectral types. The continuous line is a
second order polynomial fit to the stellar sequence. The
insets show the residuals between the galaxy and stellar polynomial FWHM values
at a given colour,
in units of 10$^{-3}$ arcsec. 
Results are shown using the default parameters for the mirror (1.2m) and
survey depth (medium)
for the colours $F1-Y$ (top left), $r-F1$ (top right) and $r-F4$ (bottom
right; F4 as the shape measuring filter).
The effect of using a larger mirror (1.5m) for $r-F1$ is also shown
(bottom left).
F1 and F4 correspond to the wavelength ranges 5500--9200\rm \AA~and
6120--8580\rm \AA~respectively.}
\end{figure*}

\subsection{Broad-band colour method\label{subsect:bb}} 

Here we investigate the possibility of
estimating the galaxy PSF from stars with the same broad-band colour.
For this to work well two conditions must be met 
($i$) there must be
a good correlation between a given broad-band colour and the PSF
size and ($ii$) this relation must be the same for both stars and
galaxies, despite their potentially differing spectra.
To assess the extent to which this applies for our fiducial instrument we have
plotted the PSF sizes for the galaxy and stellar populations against
two different colours: $F1-Y$, which can be fully
determined by an instrument containing a single optical filter
plus the $Y$ infra-red band, such as the proposed Euclid design;
and $r-F1$, which requires  $r$-band observations which could come
from the ground. The results are shown in
the top two panels of  Fig.~\ref{fig:FWHM-colour}.

It is clear from Fig.~\ref{fig:FWHM-colour} that there is a positive
correlation between the PSF FWHM and the colour i.e. redder objects tend
to be larger than bluer ones. This is unsurprising given that the
contribution to the PSF size from the diffraction limit of the telescope
is the dominant component. It is also clear that the correlation between
the PSF size and the $r-F1$ colour is tighter than with the $F1-Y$
colour. We expect this to occur because the $r-F1$ colour constrains
the slope of the SED within the lensing measuring filter F1, whereas the
$F1-Y$ colour constrains the SED slope at redder wavelengths, which are
less relevant. This interpretation is strengthened by the fact that other
colours such as $r-z$, $g-z$ and $g-Y$, which are also able to constrain
the slope of the SED over the shape measuring filter, produce correlations
almost as tight as those for $r-F1$.

For the distribution of stellar SEDs used in this paper, we can see from
Fig.~\ref{fig:FWHM-colour} that the stars occupy a well defined locus in the
FWHM versus colour space. We fit a parabola to this data and by doing so we
can estimate the PSF size from a colour alone. From the figure we see that
this line fits well through the stellar data points and thus we meet
condition ($i$) for the stars. It is fortunate that the wavelength-dependence of the PSF tends to be, to a first approximation, relatively
stable. This means that it is viable to use data from several high
signal-to-noise stars from several different fields to empirically find this
locus (provided that the wavelength independent effects on the PSF are
properly dealt with). To test condition ($ii$) we have to assess how well
the PSF size-wavelength-dependence for galaxies follows that of the stars.
We therefore calculate the residuals between the FWHM of the galaxies and
the fit to the stellar FWHM-colour relation (shown in the insets of the
figure).

The bias on the PSF size, $\langle \delta F_{\rm PSF} \rangle$,
is reduced dramatically
by including ground-based photometry, decreasing
in magnitude from
$3.6\times10^{-3}$ arcsec for $F1-Y$ to $0.16\times10^{-3}$ arcsec
for $r-F1$~\footnote{Unless otherwise stated, the photometry depths
are for a `medium-depth' survey, corresponding to a PS2 
survey (see Table~\ref{table:gbphots} for depths in the
$g,r,i,z$ and $y$ bands).}.
For a simple calculation in which the dependence of the
multiplicative bias on redshift is not taken into account,
we find from Eq.~\ref{eq:deltaF_to_m} that $\delta F_{\rm PSF}$
must be $<0.2\times10^{-3}$ arcsec for
a typical PSF size of $F=0.2$ arcsec.
The inclusion of ground-based photometry thus allows us to
meet our simple `back-of-the-envelope' requirement (Eq.~\ref{eq:deltaF_to_m}).
Using a shallower $r$-band photometry depth (PS1-like) increases
the bias by a factor of 2.5 from the value with the fiducial (medium)
depth. However, using a deeper ground-based photometry depth
(PS4-like) has no effect on the bias. This suggests that
there is an intrinsic difference between the
distribution of PSF sizes for
typical star and galaxy SEDs (measured in the $F1$ band)
with the same $r-F1$ colour, which is not reduced by
increasing the photometry depth beyond `medium'.

The wavelength-dependence of the PSF size can be reduced by increasing the size
of the primary mirror (since the telescope is diffraction-limited). We find that
the bias is reduced by a factor of 1.6 by increasing the mirror diameter from
the fiducial value (1.2m) to 1.5m (Fig.~\ref{fig:FWHM-colour} bottom-left
panel)\footnote{For these simulations the changes in the
instrumental configuration do not change the signal-to-noise of the
observations, thus the same galaxies are observed in all cases.}.
For instance, for a space telescope, such as the HST, with a 2.4m mirror, the diffraction component will contribute the same as the
CCD MTF and the jitter at $\sim7500$\AA. In this case the overall optics
becomes much less chromatic, in particular for bluer wavelengths where the CCD MTF partially compensates the diffraction effect.
This is directly reflected in the bias we estimate as $\langle \delta F_{\rm PSF} \rangle$ gets to be as small as $0.03\times10^{-3}$ arcsec (less than 5 times the value of our default configuration). On the other hand a
mirror as small as 0.6m will induce a bias in the PSF size
of $0.46 \times 10^{-3}$ arcsec, or 2.5 times larger than the bias we
obtained with the fiducial configuration.

Another way to reduce the chromaticity of the system is by using a shape
measurment filter F4 whose passband is 2/3 narrower than F1 (See
Fig.~\ref{coldep}).By using the narrower shape measurement filter we increased
to absoute bias value from +0.16 to -0.18 miliarcsec. As we can see in
Fig.~\ref{fig:FWHM-colour} (bottom-right panel), this configuration indeed
decreases the slope of the PSF FWHM--colour relation for stars and galaxies.
However, for this particular  colour of choice ($r-F4$), the `matching-up' of
stars and galaxies becomes poorer. If, for instance, we use the colour $g-z$
instead  we get a bias of -0.11, that captures the improvement we expected by
using a narrower filter.

\subsubsection{Redshift dependence}

Galaxies with the same intrinsic spectra will have different observed colours as
a result of the range of galaxy redshifts. The bias on the PSF FWHM is thus
redshift-dependent, and could potentially be very damaging to weak-lensing
tomography. In the previous section we calculated the mean bias on the PSF FWHM
for the galaxy SEDs averaged over all redshifts included in the catalogue
($z<2$). In this section we investigate the effect of the redshift dependence of
the galaxy colours.

\begin{figure}
\centerline{ \includegraphics[width=1.0\columnwidth]{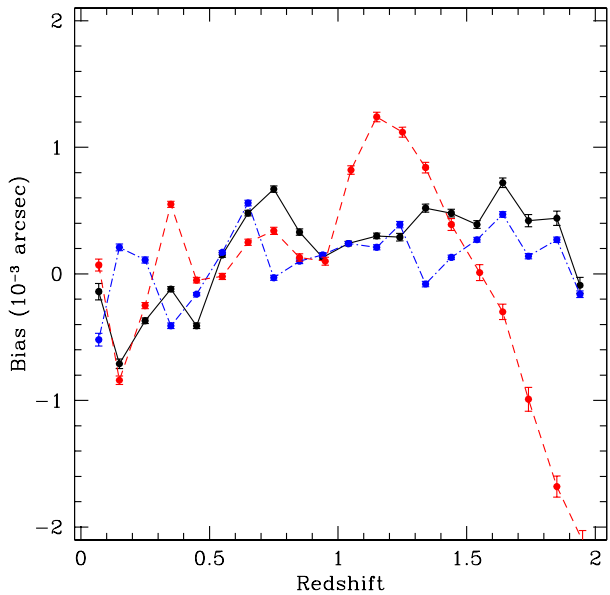}}
\caption{\label{fig:zbias}
Residual between the actual and the best estimate galaxy PSF
FWHM as a function of redshift
with the broad-band method using the $r-F1$ colour (red dashed) and the template-fitting
method using the shape measurement filters $F1$(5500--9200\rm \AA) (black solid)
and $F4$ (6120--8580\rm \AA) (blue dot-dashed).}
\end{figure}

Fig~\ref{fig:zbias} shows the redshift dependence of $\delta
F_{\rm PSF}$ for the fiducial scenario (dashed red line).
There is a large negative bias at high redshift which will affect a small
fraction of galaxies. We also see clearly the biggest limtation of our
global average method, since in that method positive and negative
contributions will cancel.

We propagate this through to dark energy related cosmological parameters as explained
in Section~\ref{sec:propagation} and divide the biases on the parameter values by the
statistical errors on the parameters found using the standard Fisher matrix approach,
for a 20,000 square degree survey with 35 galaxies per square arcminute and a total
uncertainty on each shear component of $\sigma_{\gamma}=0.35$. The results are shown
in Table~\ref{table:cospar_biases}.

\begin{table*}
\caption{\label{table:cospar_biases} Biases on dark energy related cosmological
parameters divided by the statistical uncertainty 
(0.035, 0.045 and 0.149 for $\Omega_{\rm DE}$, $w_0$ and $w_a$
respectively) on the cosmological
parameter. Results are shown for each correction method.}
\begin{center}
\begin{tabular}{|lccl|ccc|}
\hline
Colour Information &  $D$ & Lensing  & Photometry & $b(\Omega_{\rm DE})/\sigma(\Omega_{\rm DE})$ & $b(w_0)/\sigma(w_0)$ & $b(w_a)/\sigma(w_a)$   \\
       & (m)&  filter & depth&	  & & \\
\hline
\hline
 &  \multicolumn{3}{l}{Broad-band colour method} \vline & &  &  \\
$r-F1$ & 1.2 & $F1$ & Medium	& -0.05  & 0.17 & -0.47  \\
$r-F1$ & 1.2 & $F1$ & Shallow	& -0.15  & 0.34 & -0.79  \\
$r-F1$ & 1.2 & $F1$ & Deep	& -0.05  & 0.17 & -0.47  \\
$r-F1$ & 1.5 & $F1$ & Medium	& -0.04  & 0.14 & -0.36  \\
$r-F4$ & 1.2 & $F4$ & Medium	& -0.10  & 0.35 & -0.49  \\
\hline
&  \multicolumn{3}{l}{Template-fitting method} \vline & & & \\
$F1,Y,J,H,g,r,i,z,y$ & 1.2 & $F1$ &  Medium &  0.03  & 0.10 & -0.43\\
$F1,Y,J,H,g,r,i,z,y$ & 1.2 & $F1$ &  Shallow&  0.12  & 0.18 & -0.72\\
$F1,Y,J,H,g,r,i,z,y$ & 1.2 & $F1$ &  Deep   & -0.02  & 0.06 & -0.24\\
$F1,Y,J,H,g,r,i,z,y$ & 1.5 & $F1$ &  Medium &  0.02  & 0.07 & -0.31\\
$F4,Y,J,H,g,r,i,z,y$ & 1.2 & $F4$ &  Medium &  0.01  & 0.07 & -0.15\\
\hline
\end{tabular}
\end{center}
\end{table*}

We see that, for all configurations all the dark energy parameter biases we get
are smaller that the expected statistical errors. In particular, for $w_a$, the
most demanding parameter those surveys are trying to determine,  the bias 
corrected by the broad-band colour method  is less than a half of the statistical
error level for our fiducial survey configuration. The other trends follow 
those already discussed  in the context of the global averaged biases
discussed in the beginning of this Section.

\subsection{Template-fitting method \label{subsect:template}}

The previous section assumes that we use a single colour to determine the
correct PSF model to use for a given galaxy. In practice we will have more
than two filters, which will be used to calculate photometric redshifts.
We therefore also consider the use of a template fitting method to predict
the PSF FWHM of a galaxy by using all the colours available. Using ANNz
\citep{ANNz} we trained and validated a neural network using approximately
one third of the simulated galaxies available, to predict the redshifts,
spectral types and reddening of each galaxy, given the multi-colour
information. With this information we can compute the SED of each object
and use a model of the PSF wavelength-dependence to predict the PSF FWHM for
this galaxy.

A telescope model and stars will be used to build this model for the
wavelength-dependence, and the accuracy of the model will depend on the
stability of the wavelength-dependence with telescope properties, and the
number of stars available to calibrate which model to use. In the case of
the Hubble Space Telescope, a PSF model taken from the telescope design is
routinely used in conjunction with calibration from any stars in the field
to assess the telescope configuration in a given observation.
Therefore in this paper we use the exact model as given in
Eq.~\ref{eq:final} and propagate the noisy and potentially biased galaxy
SED estimates through to PSF biases and cosmological parameter biases.

\begin{figure}
\centerline{ \includegraphics[width=1.0\columnwidth]{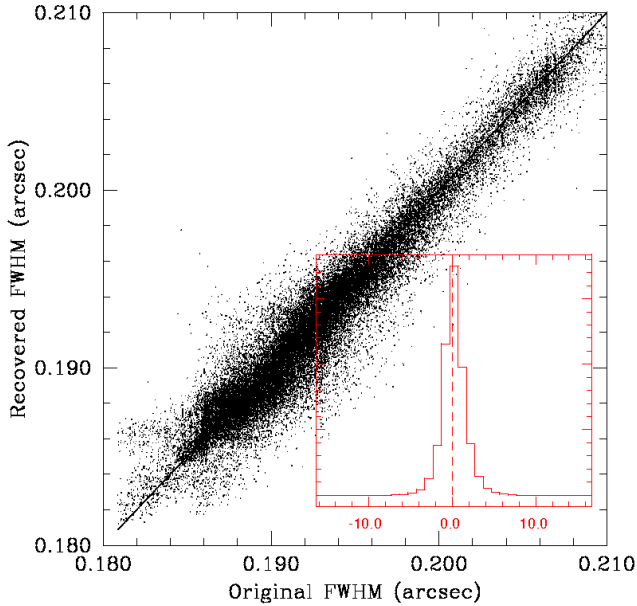}}
\caption{\label{templatefit} Comparison between the actual PSF sizes for
objects with the SED of the galaxies of the mock catalogue
and calculated PSF sizes given redshifts, reddening and spectral types
obtained through ANNz.}
\end{figure}

The comparison between this predicted PSF FWHM and the truth for the exact
galaxy SED and redshift can be seen in Figure \ref{templatefit}. We can also
compare the global averaged bias to compare  with the results we got from the
broad-band method and understadn the trends. By using our standard configuration
we obtain a bias of $\langle \delta F_{\rm PSF} \rangle =+0.19$ miliarcsec,
which is similar from what we got by using one single colour and also met our
back-of-envelope requirment (See Fig.~\ref{templatefit}). The use of  deeper
photometry, a larger 1.5m mirror or the narrower lensing filter F4 reduced
$\langle \delta F_{\rm PSF} \rangle$ to +0.18, +0.13 and +0.12 respectively,
following the expected trends.

The redshift dependence of the PSF FWHM bias is shown for the fiducial
scenario (solid black line) and for a senario with the lensing filter F4
(dot-dashed blue line) in Fig.~\ref{fig:zbias}. Both cases
show much less redshift evolution than the broad-band colour method, in
particular for $z>1.0$.

\begin{figure}
\centerline{ \includegraphics[width=1.0\columnwidth]{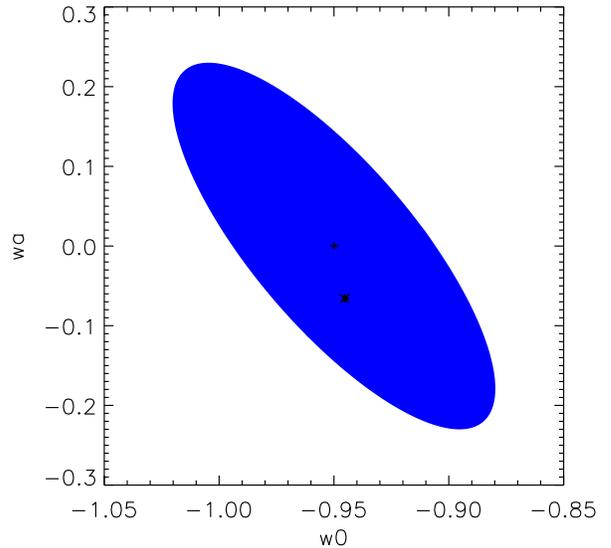}}
\caption{\label{fig:errors} Bias in the dark energy
equation of state parameters.
The ellipse shows the region of the $w_0-w_a$ space contained in the
68\% confidence level contour for a weak-lensing Euclid-type survey, where the
central values are respectively 0.0 and -0.95 (marked by a cross). The
asterisk shows the effect of the residual bias induced by the wavelength-dependence effect. }
\end{figure}

We propagate this (and all the other scenarios) through into biases on
cosmological parameters and find the results given in the second line of
Table~\ref{table:cospar_biases}. All the biases are  smaller than the
statistical errors. In Fig.~\ref{fig:errors} we show the statistical confidence level uncertainties in the space $w_0-w_a$ 
for our fiducial scenario. It can be seen in the figure that the residual bias after the correction of the wavelength-dependence is well within the 68\% CL.
In comparison to the  broad-band colour method the
template fitting method produces smaller biases on cosmological parameters
and takes more advantage of the deeper photometry.
In this case by using the deepest photometry (PS4-like)
we could see actual improvement when compared to
the default, medium depth (PS2/DES-like).

\subsection{Requirements on a simple wavelength-dependence model}
\label{sec:requirements_on_model}

\begin{figure}
\centerline{ \includegraphics[width=1.0\columnwidth]{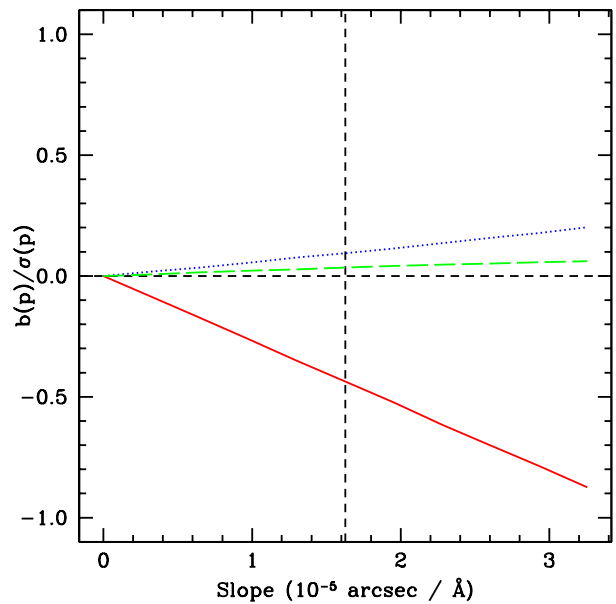}}
\caption{\label{fig:bias_plot}
Variation of the systematic to statistical
error ratio ($b(p)/\sigma(p)$) of dark energy parameters as a function of the
slope of the PSF FWHM wavelength relation. 
The dashed (green), dot-dashed (blue) 
and solid (red) lines represent the error ratio for the 
parameters $\Omega_{\rm DE}$, $w_0$ and $w_a$,
respectively.}
\end{figure}

Now we consider requirements on the parameters of a simple model for
PSF FWHM wavelength-dependence
for the template fitting method. 
We approximate the FWHM-wavelength relation to be a simple linear function. The fiducial configuration is most closely
approximated by a straight line whose slope is equal to
$1.63\times10^{-5}$ arcsec/\AA\xspace with FWHM(7350\AA) = 0.192".
We consider a range of slopes, from a pure wavelength independent
PSF to a relation twice as steep as the fiducial configuration. In
all the cases we kept the PSF FWHM constant at $\lambda=7350$\AA,
which are the central wavelenth of the lensing filters we are considering.
In Figure \ref{fig:bias_plot} we show the variation of the
systematic to statistical error for the dark energy parameters
$\Omega_{DE}$, $w_0$ and $w_a$ for the case where the
wavelength-dependence of the PSF has been corrected using the
template-fitting method.

In this Figure, as expected, one can see that the bias on cosmological
parameters increases as the wavelength-dependence of the PSF gets
stronger. For all dak energy parameters the biases lie within the region
where the statistical errors are larger than the systematic ones
$|b(p)/\sigma(p)|<1$. For the evolution of the dark energy equation of
state parameter the normalised bias $|b(w_a)/\sigma(w_a)|$ approaches
unity when the slope is double the value for the fiducial telescope mirror
size and filter width.

\section{Discussion}
\label{sec:discussion}

The wavelength-dependence of the point spread function is an
effect that has to be carefully considered for the next
generation of cosmic shear experiments, particularly if wide
bands are used for imaging. The different spectral energy
distributions of stars and galaxies means that the point spread
function obtained from stars is not the same as that for the
galaxies and this can lead to a non-negligible bias in shear
measurements.

We have, for the first time, set out a formalism for testing
the wavelength-dependence of the point spread function
for diffraction limited imaging, using the parameters
of a  Euclid-like survey. Given these characteristics the dominant
wavelength-dependence comes from the PSF size, which we parameterise
by the PSF full-width at half maximum intensity.
We have shown that 
the fractional difference
in PSF FWHM between the stars and the galaxies must be smaller
than $1\times 10^{-3}$. We find the same fractional requirement
on the PSF ellipticity difference. We have illustrated the
formalism on a fiducial Euclid-like telescope for which only
the PSF size is significantly wavelength dependent.

We investigated two different methods for correcting the effect and
found that although they give similar results for the average PSF
error, the different dependencies of the PSF size bias on redshift
leads to very different implications for biases on cosmological
parameters. For this type of analysis it is therefore necessary to
take into account the redshifting of galaxy spectra and the
cosmological parameters of interest.

The first correction method we consider matches stars of a given
colour to galaxies of the same colour. This is not expected to be a
perfect correction method because the spectral energy distributions
of two objects with the same colour are different. The stars and
galaxies need to be matched up so they have similar SEDs within the
imaging band used for cosmic shear galaxy shape measurement. This
is best achieved if the colour considered matches well to the
imaging band. We find that a telescope with a wide optical band
plus infra-red bands cannot sufficiently self-correct for PSF
wavelength-dependence using the optical minus infra-red colour
($F1-Y$). This is expected because the difference in luminosity
between bands so widely spaced is not well correlated with the
variation in luminosity within the optical band itself. The PSF
FWHM of the stars is on average $3.6\times10^{-3}$ arcsec and
a fractional error of $18\times 10^{-3}$, much greater than our
requirement ($1\times 10^{-3}$) .

To realise the full potential of cosmic shear, all planned surveys
will estimate the galaxy redshift using photometric redshifts. This
places stringent requirements on having additional photometry in
multiple wavebands, which may be obtained from the ground or from
space. This provides the ideal input into correction for the
wavelength-dependence of the PSF. We consider the colour r$-$F1,
which provides much more useful information about the spectral
energy distribution in the F1 optical band. The average difference
in PSF FWHM between the stars and galaxies now meets our
requirement. The redshift dependence shows some evolution and tends to
present  larger values for redshifts larger than 1.0. However we find that the
dark energy equation of state evolution parameter will not be  biased by more
than the statistical error if this method is used, even in the case where we use
a shallow ground based photometry.

We therefore consider an additional method in which the full range of available
wavebands are used, to match those used in photometric redshift estimation. We
take advantage of the fact that many methods for estimating the galaxy
photometric redshift also provide an estimate of the galaxy spectral energy
distribution. This allows a full model of the galaxy spectrum to be used in
correcting for the PSF wavelength-dependence. The extent to which this is
helpful depends on how well the instrument PSF wavelength-dependence is already
known. We do already have a reasonable model for the wavelength-dependence.
Additionally we expect the wavelength dependence to be quite stable as a
function of time, and furthermore it should be very well calibratable using
stars. In this work we assume that the model for the wavelength-dependence is
therefore very well known, and the limiting factor in the analysis comes from
uncertainties in the galaxy photometry, which limit our knowledge of the galaxy
redshift and spectral energy distribution. We find for the several cases we
tested that the redshift dependence of the bias is smaller, when compared to the
broad-band method and, as a consequence, the dark energy cosmological parameter
biases are also smaller.

We consider the effect of a different telescope mirror size and
imaging filter width. We find that the large mirror size does
decrease the biases, as expected due to the smaller contribution to
the wavelength-dependence from the diffraction limit. The reduction
is around 20 per cent for both correction methods. Using a narrower
filter reduces the scatter on the PSF FWHM but does not decrease the
bias, for our particular configuration.

Finally we consider a general linear PSF wavelength-dependence which matches
well to the Euclid-like fiducial configuration. This allows a requirement on the
linear slope of the PSF to be obtained for a given required accuracy on
cosmological parameters. We find that, by using the template fitting method, a
survey in which the  wavelength-dependence of the PSF is  twice as strong as
our  fiducial Euclid-like survey just meets the requirement that
$|b(p)/\sigma(p)|<1$.

In this work we have studied the first order effects of a wavelength  
dependent PSF and have shown that this effect can be mitigated with  
the addition of photometric data. The next step is to consider  
higher order effects such as colour gradients and the spatial  
correlation function of galaxy colours. These issues will be tackled  
and discussed in later work.

\section*{Acknowledgements.}
\bsp

We are grateful to Peter Capak for providing the code used to generate the mock
photometric catalogue. We thank Mark Cropper, Jerome Amiaux, Peter Doel, Ofer
Lahav, James Kingston, Steve Kent, Michelle Antonik, Stephane Paulin-Henriksson,
Gary Bernstein, Tom Kitching, Alexie Leauthaud and Gary Bernstein for helpful conversations. ESC
acknowledges support from FAPESP (process number 2009/07154-8). LMV acknowledges
support from STFC. SLB and FBA thanks the Royal Society for support in the form
of a University Research Fellowship. The research described in this paper was
performed in part at the Jet Propulsion Laboratory, California Institute of
Technology, under a contract with the National Aeronautics and Space
Administration.

\bibliographystyle{mn2e}
\bibliography{bibfile}


\appendix

\section{Multiplicative and additive shear measurement errors from PSF mis-estimates}
\label{sec:mc_quadmoms}

We follow~\cite{paulinavrb08}, who define object size $R$ and two-component ellipticity $\ellipticity$ in terms of unweighted quadrupole moments $Q_{ij}$ as
\begin{eqnarray}
\label{eq:sizedef}
R^2 & = & Q_{11} + Q_{22}\\
\label{eq:ellipticitydef}
\ellipticity_1 & = & \frac{Q_{11}-Q_{22}}{Q_{11} + Q_{22}}\\
\ellipticity_2 & = & \frac{2Q_{12}}{Q_{11} + Q_{22}}.
\end{eqnarray}
It is shown in \cite{paulinavrb08} that the systematic bias on a galaxy ellipticity
component $\delta\bs{\ellipticity}^{\rm sys}_{{\rm gal}\, i}$ can be approximated in
terms of the mis-estimates of the PSF size
$\delta R_{\rm PSF} $ and PSF ellipticity $\delta
\bs{\ellipticity}_{{\rm PSF}\, i}$ as
\begin{equation}
\label{eq:deltaesys}
\delta\bs{\ellipticity}^{\rm sys}_{{\rm gal}\, i} \approx
\left(\frac{\size_{\rm PSF}}{\size_{\rm gal}}\right)^2
\left(
2 \left( \bs{\ellipticity}_{{\rm gal}\, i} -
\bs{\ellipticity}_{{\rm PSF}\, i} \right)
 \frac{ \delta R_{\rm PSF} }{\size_{\rm PSF}}
 -  \delta \bs{\ellipticity}_{{\rm PSF}\, i}
 \right)
\end{equation}
where $\bs{\ellipticity}_{{\rm gal}\, i}$ are the original (pre-PSF but post shear) galaxy ellipticity components and $R_{\rm gal}$ is the galaxy size. Similar definitions apply for the PSF ellipticity and size.
This assumes the PSF and galaxy size and ellipticity measurements are
made using unweighted quadrupole moments.
This propagates into a bias on the shear estimate $\hat{\gamma}_i = \gamma_i + \delta \gamma_i$ as
\be
\delta \gamma_i = \frac{\delta \bs{\ellipticity}_{{\rm gal}\, i}^{\rm sys}}{P^{\gamma}}
\label{eq:Pgammaeq}
\ee
where $P^\gamma$ is the shear responsivity given by $P^\gamma = 2-\langle \left|\bs{\ellipticity}\right|^2 \rangle \sim 1.8 $ for simple shear measurement methods.
Similarly $\gamma_i=\ellipticity_{{\rm gal}\, i}/P^{\gamma}$.


In
this paper we quantify object sizes in terms of the slightly more intuitive quantity, the
Full-Width at Half-Maximum (FWHM), $F$ where
\begin{equation}
F = 2 \sqrt{\ln 2} \,\, R
\label{eq:fwhm}
\end{equation}
for a Gaussian profile, which we use for the PSF component in this paper.
We combine Eq.~\ref{eq:mcdef}, Eq.~\ref{eq:fwhm} and Eq.~\ref{eq:Pgammaeq} and compare with Eq.~\ref{eq:deltaesys}
to find
\bea
m &= &
2
\left(\frac{R_{\rm PSF}}{R_{\rm gal}}\right)^2
\,\frac{\delta F_{\rm PSF}}{F_{\rm PSF}}
 \label{eq:m}
\\
\bs{c_i} &=&
-
\frac{1}{P_{\gamma}} 
\left(\frac{R_{\rm PSF}}{R_{\rm gal}}\right)^2
\left(
2\, \bs{\ellipticity}_{{\rm PSF}\, i}
\, \frac{\delta F_{PSF}}{F_{\rm PSF}}
 + \delta \bs{\ellipticity}_{{\rm PSF}\, i}
 \right)
 \label{eq:c}
\eea where the second term in $c_i$ dominates for typical future
surveys because ${\ellipticity}_{{\rm PSF}\, i}$ is small, in
addition to the small value of the fractional uncertainty in the PSF
FWHM. We have ignored the contribution to $c_i$ from intrinsic
galaxy ellipticities, as appropriate for the case where the average
$c_i$ over randomly oriented galaxy ellipticities is required. Where
useful we have converted PSF sizes into FWHM values but we have left
galaxy sizes written using $R$ since the conversion from $R$ to $F$
is dependent on object profile, and is far from Gaussian for
galaxies. As in \cite{paulinavrb08} we assume $R_{\rm gal} \ge 1.5
R_{\rm PSF}$ and thus $R_{\rm PSF}/R_{\rm gal}\simeq 0.5$. Thus, as
expected, multiplicative errors arise from errors in the PSF size
and additive errors arise from errors in the PSF ellipticity.



\label{lastpage}
\end{document}